\documentclass[conference]{IEEEtran}
\IEEEoverridecommandlockouts
\usepackage{cite}
\usepackage{amsmath,amssymb,amsfonts}
\usepackage{algorithmic}
\usepackage{graphicx}
\usepackage{textcomp}
\usepackage{xcolor}
\usepackage{siunitx}
\usepackage{booktabs}
\usepackage{multirow}

\begin{document}
\title{Internet of Buoys: An Internet of Things Implementation at Sea
\thanks{This work has been accepted by IEEE Asilomar 2020.}
}


\author{\IEEEauthorblockN{Michiel Sandra\IEEEauthorrefmark{1}, Sara Gunnarsson\IEEEauthorrefmark{1}\IEEEauthorrefmark{2} and
		Anders J Johansson\IEEEauthorrefmark{1}}
	\IEEEauthorblockA{\IEEEauthorrefmark{1}Department of Elecrical and Information Technology, Lund University, Sweden}
	\IEEEauthorblockA{\IEEEauthorrefmark{2}Department of Electrical Engineering, KU Leuven, Belgium}
	\IEEEauthorblockA{Email: \{michiel.sandra, sara.gunnarsson, anders\_j.johansson\}@eit.lth.se}}

\maketitle

\begin{abstract}
Internet of Things (IoT) applications are emerging in many different areas, including maritime environments.
One of the applications in this area is the monitoring of buoys at sea.
To realize wireless tracking of buoys, an accurate prediction of the path loss in an open-sea environment is essential. So far, channel measurements at sea have mainly been conducted with antennas placed a couple of meters above the sea surface, which is higher than the buoys themselves.
Therefore, we investigated the validity of the published channel models at sea by means of path loss measurements using a LoRa
link 
with a transmitter antenna height of 0.35\,m and a base station antenna height of 2.65\,m and 5.2\,m. 
Our results show that the round earth loss model is not accurate at these antenna heights. 
The ITU-R P.2001-3 model and a model by Bullington 
show a better agreement with our measurements. 
However, the difference between our two measurement campaigns shows that more investigation is needed on the dependence of the path loss on the sea state.
Additionally, the availability of Sigfox, Narrowband Internet of Things (NB-IoT), and The Things Network at sea has been explored.
We found that NB-IoT and Sigfox can be used for IoT applications in the tested area at low antenna heights.
\end{abstract}


\section{Introduction}
Recent innovations in wireless technology have opened up for the deployment of Internet of Things (IoT) applications in diverse areas, such as maritime environments.
One of the applications within this area is the monitoring of buoys at sea. Buoys can have different functionalities, such as navigation and hazard marking. In these cases, buoys are an important complement to Global Navigation Satellite System (GNSS) and sea charts to maintain safety at sea. However, buoys disappear due to ice, boats colliding with them or due to bad weather conditions. 
Organisations that are responsible for the maintenance of buoys at sea, such as harbours and maritime departments, have an interest in embedding electronics in all their  buoys, such that these can autonomously report their location. 
The embedded system has to be self-contained and must preferably last as long as the buoy
itself because changing the batteries of a buoy at sea is an expensive operation. 
Also, solar power is not feasible for all types of buoys due to practical constraints.
The location of buoys can be obtained using GNSS; nevertheless, localization techniques like triangulation based on direction of arrival (DoA) estimation between the buoys could be a more energy efficient and cost effective alternative.
In addition to the localization, buoys could be equipped with all kinds of sensors that can be useful for applications like scientific research and traffic monitoring.

Evidently, buoys have to be connected to a wireless network to report their location. Satellite networks
and General Packet Radio Service (GPRS) networks have proven useful for applications with buoys \cite{kazdaridis_buoy_energy}\cite{delpizzo_buoy_gprs}. Nonetheless, low-power wide-area networks (LPWANs) like Sigfox, The Things Network (TTN) and Narrowband Internet of Things (NB-IoT) are more energy efficient, making them suitable for the application. Note that the physical layer of TTN is LoRa.

Despite the increasing popularity of LPWANs, the availability of the networks is often limited to certain areas. To tackle this issue, a custom LPWAN for IoT applications at sea could be deployed. Additionally, the coverage of LPWANs can be expanded by implementing an ad-hoc network when buoys are too far from the base stations on shore.
Both solutions require a channel model that accurately describes the relation between the path loss and system parameters like antenna height. Since the antenna height on some kinds of buoys is lower than \SI{3}{m}, special attention should be paid to the validity of channel models at low antenna heights. So far, channel models have only been validated by measurements with the antennas placed a couple of meters above the sea surface \cite{yang_rel_model_sea}\cite{wang_broadband_sea_meas_scatter}. For this reason, we
have measured the path loss at sea with our custom-built setup at low antenna heights using a LoRa point-to-point (P2P) link, and compared the results with published path loss models to check their validity. Similar measurements have been performed in \cite{petajajarvi_lora_coverage}\cite{parri_lpwan_sea}. However, only the free space and log-distance path loss model are considered in \cite{petajajarvi_lora_coverage}, and only the free space path loss model in \cite{parri_lpwan_sea}.
Besides the path loss measurements, we explored the availability of TTN, NB-IoT and Sigfox at sea.

This paper is structured as follows. In Section~\ref{sec:modeling}, we introduce the published path loss models for over-the-sea radio wave propagation as later used for comparison. Section~\ref{sec:setup} describes our measurement setup and campaigns, and Section~\ref{sec:res} presents our results. Finally, we conclude the paper in Section~\ref{sec:concl}.

\section{Path loss modeling at sea \label{sec:modeling}}
\begin{figure}[bt!]
    \centering
    \includegraphics[width=0.8\linewidth]{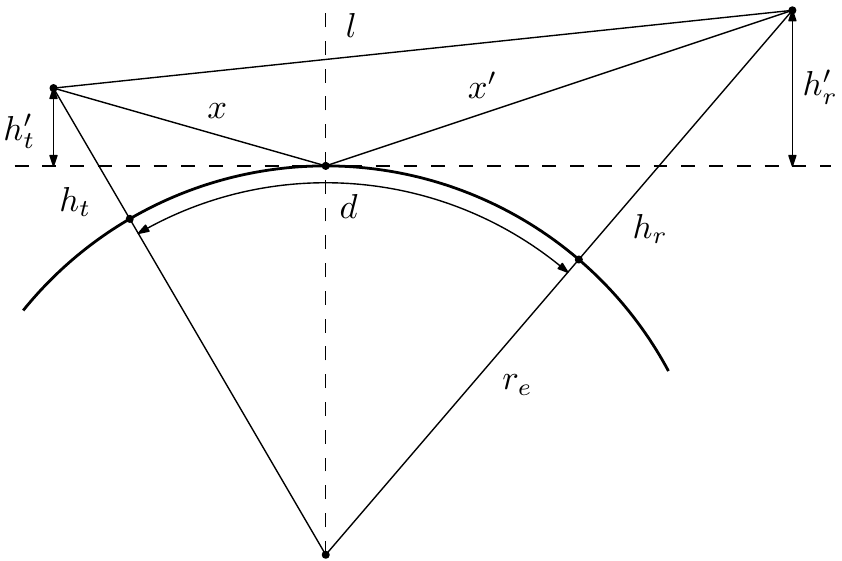}
    \caption{The two-ray model applied in the round earth geometry. $h_t$ and $h_r$ are the antenna heights, $d$ is the great-circle distance between transmitter (Tx) and receiver (Rx), $r_e$ is the earth's radius, $l$ is the length of the direct ray between Tx and Rx, $x$ is the distance between the Tx antenna and point of reflection, and $x'$ is the distance between the point of reflection and the Rx antenna. $h_t'$ and $h_r'$ are the antenna height relative to the the imaginary plane tangent to the earth at the point of reflection. \label{fig:round_earth}}
  \end{figure}
The sea is a challenging environment for wireless communication because
its surface gives rise to reflections that will
interfere with the direct ray at the receiver.
In the general case, these reflected waves consist of a specular and a diffuse component \cite{karasawa_mulitpath_sea_old}. 
In an environment where 
a single ground reflection is dominating the multipath effect,
such as the sea, a two-ray model is often used. This neglects the diffuse component, and is expressed as,
\begin{equation}
    L_p = \left(\dfrac{\lambda}{4 \pi}\right)^{-2} \left|\dfrac{1}{l}+R\dfrac{1}{x+x'} \exp\left(j\,\dfrac{2\pi(x+x'-l)}{\lambda}\right)\right|^{-2}, \label{eq:tworay}
\end{equation}
where $\lambda$ is the radio frequency (RF) wavelength and $R$ the reflection coefficient. All other variables can be derived from the geometry in Fig.~\ref{fig:round_earth} where the curvature of the earth is taken into account. Note that in \eqref{eq:tworay} the antenna gain is assumed to be the same in the direction of the direct ray and the reflected ray.

The two-ray model is characterized by an oscillation around the free-space path loss up to the critical distance, which is approximately equal to,
\begin{equation}
    d_c \approx \dfrac{4 h_t h_r}{\lambda},
\end{equation}
where $h_t$ is the Tx antenna height, and $h_r$ is the Rx antenna height.
For distances larger than $d_{c}$, diffraction due to the earth's curvature needs to be considered \cite{itur_diffraction}, which is also the propagation mechanism that makes beyond-the-horizon communication possible. 
Beyond the horizon, $d$ is greater than the great-circle distance $d_h$ where the imaginary line between the transmitter (Tx) and receiver (Rx) antenna touches the earth's surface. This distance is calculated as,
\begin{equation}
    d_h = r_e \left( \arccos\left(\dfrac{r_e}{r_e + h_t}\right) +  \arccos\left(\dfrac{r_e}{r_e + h_r}\right) \right),
\end{equation}
where $r_e$ is equal to the radius of the earth.
Published models have presented different ways to incorporate diffraction loss.
In the following paragraphs, we introduce the models that we have used to compare our measurement results to. 

The round earth loss (REL) model applies the two-ray model in a round earth geometry and takes the surface roughness and curvature of the earth into account by adapting the reflection coefficient $R$ in \eqref{eq:tworay} and multiplying \eqref{eq:tworay} with a diffraction loss factor. Two variables are used to model the surface roughness: $\beta_0$, the root mean square of the surface slope, and $\sigma_h$, the standard deviation of the surface slope. The REL model makes use of a model for the diffraction loss by Bullington \cite{bullington_diff_old}, which is valid for distances beyond the horizon. However, the authors of the REL model proposed a method to also use this diffraction model for distances between $d_{60}$ and $d_h$. The distance $d_{60}$ is the great-circle distance between Tx and Rx where the first Fresnel zone is 60\,\% clear, and is expressed as,
\begin{equation}
    d_{60} = \dfrac{1.5949 \cdot 10^{-10} f h_t h_r \left(\sqrt{h_t}+\sqrt{h_r}\right)}{3.89\cdot 10^{-11} f h_t h_r + 4.1 \left(\sqrt{h_t}+\sqrt{h_r}\right)},
\end{equation}
where $f$ is the center frequency. 

The ITU Recommendation P.2001-3 is a general purpose wide-range terrestrial propagation model in the frequency range \SI{30}{MHz} to \SI{50}{GHz} \cite{itur_general}. In addition to the normal propagation close to the surface of the earth, the model considers anomalous propagation, tropospheric scatter and ionospheric propagation. Nevertheless, the three last mentioned propagation mechanisms are negligible for the frequency band used in our measurement campaigns. A unique property of this model is its ability to predict the path loss that is not exceeded for a given percentage $T_{pc}$ of an average year. Although this gives an estimate of the fading during one year, the path loss can not be calculated for a specific sea state.

In addition to the model for the diffraction loss used in the REL model, Bullington \cite{bullington_diff_old} also included a convenient method to predict the path loss considering the curvature of the earth. The path loss is estimated by calculating the path loss using a two-ray model assuming the earth is planar, and then correcting this by adding a loss given by Bullington \cite{bullington_diff_old}.
This method is valid as long as the antenna height is below a certain value. For \SI{868}{MHz}, the antenna height has to be lower than \SI{15}{m}.
This model does not take the roughness of the sea into account, as the earth's surface is considered to be a smooth perfect conductor ($R=-1$).

The log-distance path loss model is an empirical model that is widely used to compare and characterize the path loss in different environments, and is given by,
\begin{equation}
    L_{p|\mathrm{dB}}(d) = 10\,n \log_{10}\left(\dfrac{d}{d_0}\right),
\end{equation}
where $L_{p,0}$ is the path loss at a reference distance $d_0$, and $n$ is the path loss exponent.
When the path loss is equal to the free-space path loss, $n$ is equal to 2. In the case of a two-ray path loss model, $n \approx 4$ for $d \gg d_c$ when the surface is smooth, planar and a perfect conductor ($R=-1$). 


\section{Measurement setups and campaigns\label{sec:setup}}
\begin{figure}[bt!]
    \begin{center}
        \includegraphics[width=0.8\columnwidth]{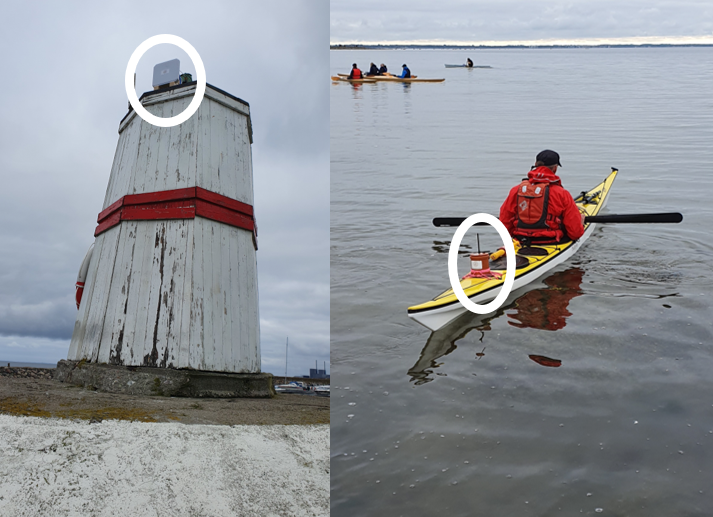}
    \end{center}
    \caption{Base station (left) and mobile measurement unit (right) during the second measurement campaign with an antenna height of \SI{5.2}{m} and \SI{0.35}{m}, respectively.\label{fig:mmubs}}
\end{figure}
\begin{figure}[bt!]
    \begin{center}
        \includegraphics[width=0.9\columnwidth]{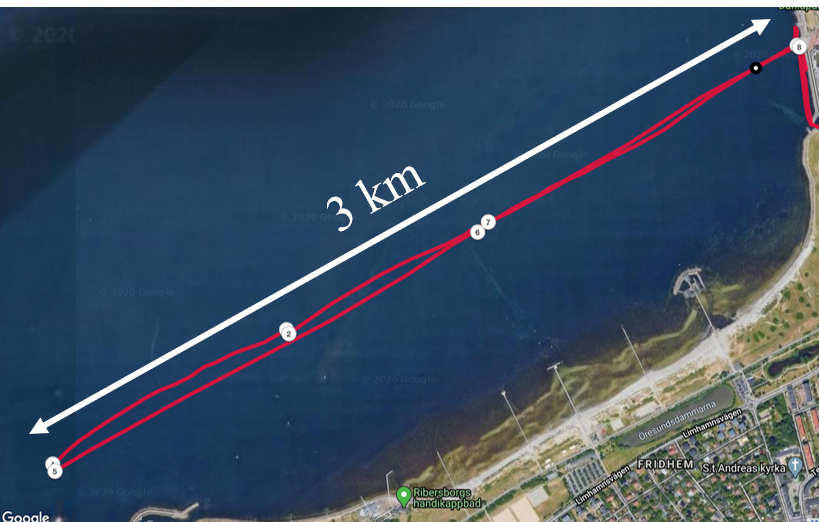}
    \end{center}
    \caption{Trajectory of the kayak in measurement 1 (red line). The start/stop location and the location of the BS is the white dot in the right upper corner of the figure. 
    \label{fig:map1}}
\end{figure}
\begin{figure}[bt!]
    \begin{center}
        \includegraphics[width=0.7\columnwidth]{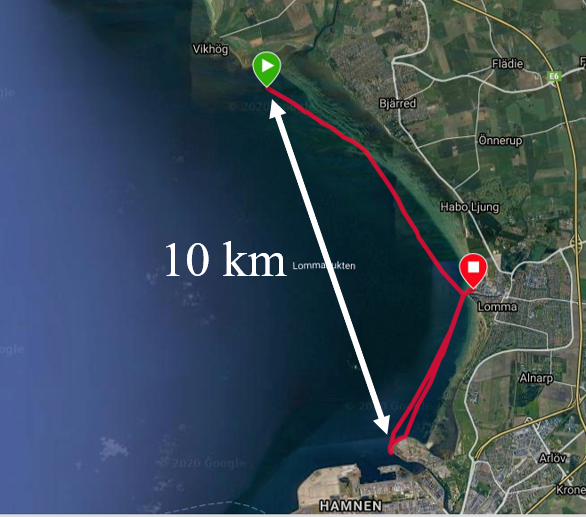}
    \end{center}
    \caption{Trajectory of the kayak in measurement 2 (red line). The green flag is the location of the BS and the start location (Vikhög). The red flag is the stop location (Lomma).
    \label{fig:map2}}
\end{figure}
We performed two separate measurement campaigns, at different places and with slightly different hardware.
The measurements were done with a mobile measurement unit (MMU) and a LoRa base station (BS), as shown in Fig.~\ref{fig:mmubs}. The MMU consists of batteries, a micro controller and three radio modules, i.e. NB-IoT (u-blox SARA-N211 \cite{saran211}), Sigfox (WSSFM10R1 \cite{wisol}) and LoRa (HopeRF RFM95W \cite{hoperf}), to connect to the corresponding networks. The modules are connected to the same antenna using an RF switch. The BS only has a LoRa radio module (HopeRF RFM95W for first campaign and Semtech SX1261MB2BAS board \cite{sx1261} for the second), and the rest of the hardware is identical.

Using LoRa for path loss measurements has a couple of benefits. First, the setup is convenient and cost-effective to build. Second, due to its compact size and low weight, our measurement setup is easy to carry and mount to different vessels. Besides, as LoRa is a low power technology, our setup is able to be active for a few weeks up to months. 
However, the setup has some drawbacks. Only the average of the received power of a LoRa message is measured, and the initial accuracy of RSSI (Received Signal Strength Indicator) is not high. To mitigate this, we calibrated our setup to minimize the errors. We used a step attenuator connected between the antenna ports of the MMU and the BS. The error of each RSSI measurement level was calculated and then compensated for in the measurement results.

\section{Results\label{sec:res}}
\begin{figure*}[ht!]
    \begin{center}
        \includegraphics[width=\textwidth]{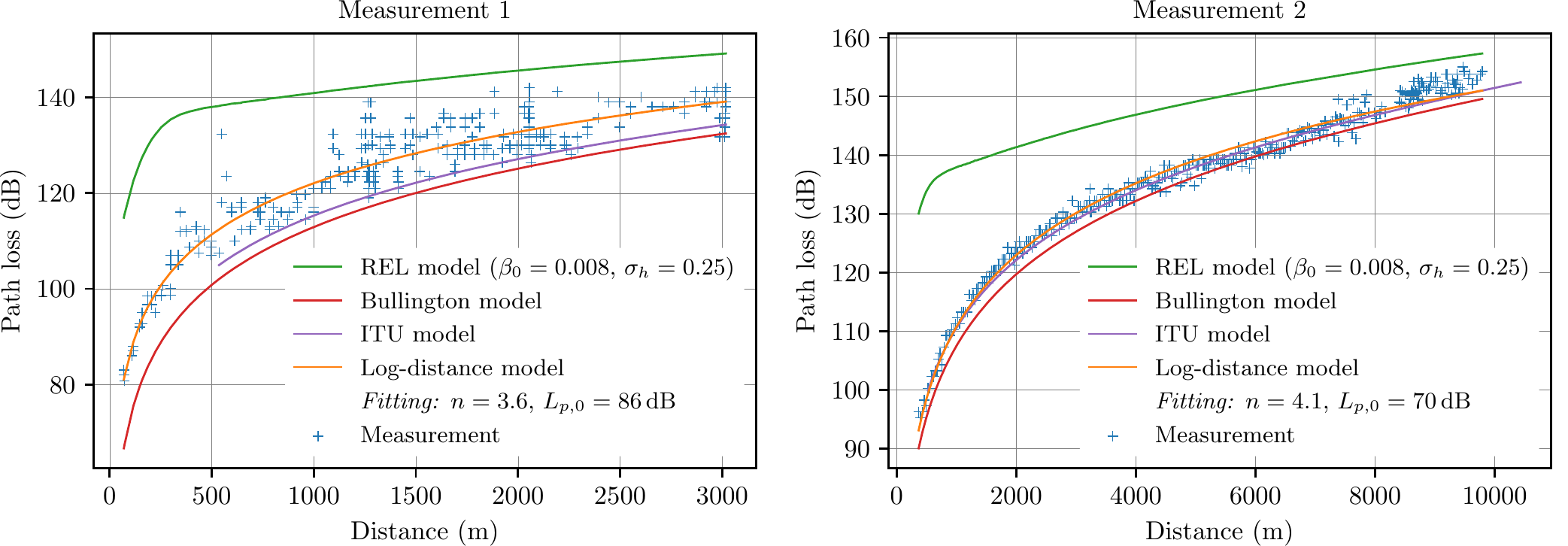}
    \end{center}
    \caption{Path loss as a function of distance\label{fig:results}}
\end{figure*}

During the campaigns, the MMU was mounted on a kayak at an antenna height of \SI{0.35}{m} using a vertical polarized 0 dBi omnidirectional antenna. The BS was placed on shore at a fixed location for the respective measurements. The first campaign was along the shore of Malmö, as shown in Fig.~\ref{fig:map1}. A maximum distance of \SI{3}{km} between the MMU and BS was obtained with the kayak, using an omnidirectional antenna at the BS, identical to the one on the kayak. The second campaign aimed at covering longer distances, and the BS LoRa module was replaced by a more accurate one. The BS antenna was exchanged for a circular polarised directional antenna with a gain of 9\,dBi directed towards the planned trajectory. The resulting polarisation loss has been compensated for in the results by subtracting the antenna gain with 3\,dB. For this campaign, the network coverage measurements were omitted, and a NB-IoT module was added to the BS to monitor the measurements. The BS was placed on a lighthouse in the harbour of Vikhög, and the MMU was mounted to a kayak. We obtained a maximum distance between the MMU and BS of \SI{10}{km}, as shown in Fig.~\ref{fig:map2}. The sea state in the second campaign was also noticeably calmer than in the first one. All the measurement parameters are summarized in Table~\ref{tab:measpar}.

Sigfox and TTN messages were sent every 10 minutes, as allowed by regulations. Since NB-IoT uses a licensed band, messages could be sent as frequent as required. For the LoRa path loss measurements, a different band was used than those used by TTN, so we could
perform more frequent measurements.
The BS receives a LoRa message from the MMU every 17 seconds and a RSSI value is captured. This value and the GPS coordinates were sent through the LPWANs to be monitored on shore. A backup was also made on an SD card in the BS.

\begin{table}[]
    \centering
    \caption{Measurement parameters}
    \renewcommand*{\arraystretch}{1.3}
    \label{tab:measpar}
        \begin{tabular}{lcc}
            \toprule
            \multicolumn{1}{l}{Parameter} & \multicolumn{1}{c}{Measurement 1} & \multicolumn{1}{c}{Measurement 2} \\ 
            \midrule
            Frequency        &     \multicolumn{2}{c}{\SI{869.5}{MHz}}  \\
            Output power & \SI{17}{dBm} & \SI{18.3}{dBm} \\
        MMU antenna height      &             \multicolumn{2}{c}{\SI{0.35}{m}} \\
        BS antenna height       &   \SI{2.65}{m}   &  \SI{5.2}{m}  \\
        MMU antenna gain        &  \multicolumn{2}{c}{\SI{0}{dBi}} \\
        BS antenna gain        & \SI{0}{dBi}      &   \SI{9}{dBi} (circ. pol.) \\
        LoRa spreading factor & \multicolumn{2}{c}{SF8}\\
        LoRa bandwidth & \multicolumn{2}{c}{\SI{10.4}{kHz}}  \\
        BS sensitivity & \multicolumn{2}{c}{\SI{-138}{dBm}} \\
        Average temperature & \SI{19.5}{\celsius} & \SI{19.9}{\celsius} \\
        Maximum distance &  \SI{3.02}{km} & \SI{9.79}{km} \\
        Number of measurements &            315          &           325    \\
            \bottomrule
        \end{tabular}
\end{table}

In Fig.~\ref{fig:results}, the path loss measurement results of both campaigns are plotted as a function of distance. The results of the first campaign are to the left, and the results of the second campaign are to the right. The path loss models that are introduced in Section~\ref{sec:modeling} are also plotted in Fig.~\ref{fig:results} for comparison. 
From the plots, it can be seen that the predicted loss by the REL model is always higher than the measured values.
On the other hand, the Bullington and the ITU model are closer to the measurements, and predict a lower path loss than the fitted log-distance model.
Besides, note the difference in spread of the measurements between the two campaigns. 

During the second measurement, breaks were taken at \SI{8}{km} and \SI{10}{km} where the MMU was on land and behind rocks, respectively. 
Because the measurement points are irrelevant in these situations, they are left out. For the calculation of the REL model, the same sea state is assumed as in \cite{yang_rel_model_sea}, as this corresponds to our assessment of the actual state. Hence, the same values for $\beta_0$ and $\sigma_h$ are used here. 
The diffraction loss for the REL model and Bullington model is calculated assuming that the ratio of the effective earth radius to the true radius is equal to one.
The log-distance model is fitted using the least-squares method. For $d_0$, we chose a value of \SI{100}{m} because this lies beyond $d_c$, which is equal to \SI{11}{m}  and \SI{21}{m} for the first and second campaign, respectively. Both measurement campaigns stayed within the horizon distance $d_h$, which is equal to \SI{7.9}{km} and \SI{10.3}{km} for the first and second campaign, respectively.

\begin{table}[]
    \centering
    \renewcommand*{\arraystretch}{1.3}
    \caption{RMSE and MAE results\label{tab:error}}
        \begin{tabular}{lrrrr}
            \toprule
            \multicolumn{1}{l}{\multirow{2}*{Model}} & \multicolumn{2}{c}{Measurement 1} & \multicolumn{2}{c}{Measurement 2} \\ 
            \cmidrule(lr){2-3} \cmidrule(lr){4-5}
            & RMSE & MAE & RMSE & MAE \\ 
            \midrule
            REL        & 19.1  & 17.4 & 14.4 & 12.5 \\
            Bullington & 9.7 & 8.7 & 3.1 & 2.7 \\
            ITU       & 6.8  & 5.2 & 1.7 & 1.3 \\
            Log-distance & 4.0  & 3.1 & 1.7 & 1.4 \\
            \bottomrule
        \end{tabular}
\end{table}

To elaborate on the validation of the models, the root mean square error (RMSE) and the mean absolute error (MAE) are calculated, and presented in Table~\ref{tab:error}. According to the RMSE and MAE values, the REL model has the biggest disagreement with our measurement results.
A possible explanation is that the method used in the REL model for predicting the diffraction loss before the horizon distance is not valid for low antenna heights.
The Bullington and ITU model are better at predicting the path loss. However, the agreements are better in the second measurement than in the first. This could be due to the calmer sea state and the higher placed BS antenna. The lower path loss exponent in the first measurement shows that the rate at which the path loss increases is smaller.
Also, note that the path loss exponent in the second campaign is approximately equal to four, which corresponds to a two-ray model with a perfectly conducting smooth surface. 

During our first measurement, almost all the messages from the Sigfox and NB-IoT were received. No messages were received from TTN. This result is not surprising, as TTN is a crowd sourced network. The coverage of this type of network relies on volunteers setting up BSs, and they probably have limited interest in providing coverage at the sea.

\section{Conclusion\label{sec:concl}}
The Internet of Buoys has provided us with new questions regarding channel modeling at sea. With our custom-built setup, we have measured the path loss at sea at low antenna heights and investigated the validity of different path loss models. Our results show that the REL model does not give accurate results at low antenna heights, and that the ITU model and Bullington model agree better with our measurement results. However, the difference between our two measurement campaigns shows that there is further investigation needed on the dependence of the path loss on the sea state. 
Additionally, the availability of the deployed low-power wide-area networks LPWANs at sea has been explored in the surroundings of Malmö. Our results led to the conclusion that the Sigfox and NB-IoT network can be used for IoT applications in the tested area at low antenna heights.

\bibliographystyle{IEEEtran}
\bibliography{references}

\end{document}